\documentclass[aps,prl,reprint,superscriptaddress]{revtex4-1}

\usepackage{graphicx}
\usepackage{epstopdf}
\usepackage{amsmath}
\usepackage[latin1]{inputenc}
\usepackage[svgnames]{xcolor}
\usepackage[geometry]{ifsym}

\begin{document}

\title{Homogeneous superconducting phase in TiN film : a complex impedance study}
\author{P. Diener}
\affiliation{SRON Netherlands Institute for Space Research, 3584 CA Utrecht, The Netherlands.}
\email[]{diener@lps.u-psud.fr}
\author{H. Schellevis}
\affiliation{Delft Institute of Microsystems and Nanoelectronics, Delft University of Technology, 2628 CD Delft, Netherlands.}
\author{J.J.A. Baselmans}
\affiliation{SRON Netherlands Institute for Space Research, 3584 CA Utrecht, The Netherlands.}

\date{\today}

\begin{abstract}

The low frequency complex impedance of a high resistivity $92\,\mu\Omega cm$ and 100 nm thick TiN superconducting film has been measured via the transmission of several high sensitivity GHz microresonators, down to $T_{C}/50$. The temperature dependence of the kinetic inductance follows closely BCS local electrodynamics, with one well defined superconducting gap. This evidences the recovery of an homogeneous superconducting phase in TiN far from the disorder and composition driven transitions. Additionally, we observe a linearity between resonator quality factor and frequency temperature changes, which can be described by a two fluid model. 

 \end{abstract}


\maketitle

Thin films of titianium nitride (TiN) have recently been proposed \cite{Leduc_APL10} for radiation detection using Kinetic Inductance Detectors \cite{Day2002}. TiN and other strongly disordered superconducting materials have a high surface impedance for radiation with a photon energy $h\nu>2\Delta$ which allows highly efficient radiation absorption. Additionally, disordered superconductors have a long magnetic penetration depth. This results in a relatively large value of the kinetic inductance, which increases the responsivity of the devices.

Recent work has revealed strong deviations from standard BCS theory for superconductors with a normal state resistivity in the range of 100 $\mu\Omega cm$ and higher \cite{Driessen2012}, clearly in violation of the Anderson theorem which states that disorder does not affect the properties of the superconducting state \cite{Anderson1959}. These deviations increase with higher disorder, leading to systems in which a superconductor insulator transition (SIT) is observed, typically when the Ioffe-Regel parameter $k_{F}\ell\sim1$. SITs have been observed in thin films of several materials by changing the disorder via a film treatment, or by varying the thickness, the film stoichiometry or the applied magnetic field \cite{Shahar_PRB92, Marrache_PRB08, Okuma1998, Bielejec_PRL02}. In TiN a SIT has been evidenced few years ago in $\sim 5 nm$ thick films by magnetic field and thickness changes \cite{Baturina2004, Hadacek_PRB04}. Several studies also report the presence of an inhomogeneous superconducting phase close to the SIT \cite{Baturina2007, Sacepe_PRL08, Baturina_PhysB05, Baturina2_PRL07, Vinokur2008}. 

One expects to recover a classical BCS superconducting phase when going to lower disorder/thicker films. In contrast, the only one spectroscopic study on a low disordered TiN film reports on the presence of a non uniform state comprising of superconducting and normal areas \cite{Escoffier_PRL04}. These results are discussed in the context of mesoscopic fluctuations close to a superconducting to normal transition. Indeed, TiN also exhibits a transition with composition: as reported recently, superconductivity disappear in the N sub stoichiometric range \cite{Leduc_APL10}. The presence of an inhomogeneous order parameter in a low disordered TiN film points out the necessity to disentangle the thickness and composition transitions. 

In this paper, we report on the complex impedance study of several microresonators made of a 100 nm thick, relatively weakly disordered TiN film, having $k_{F}\ell=12.7$ and a resistivity of $92\,\mu\Omega cm$. The film has been characterized extensively, and the resonators have internal quality factors up to $10^{7}$. This allows an accurate determination of the superconducting gap from the temperature resonance frequency shift which is directly proportional to the kinetic inductance or superfluid density changes. 

\begin{figure}[t]
		\includegraphics[width=1\linewidth]{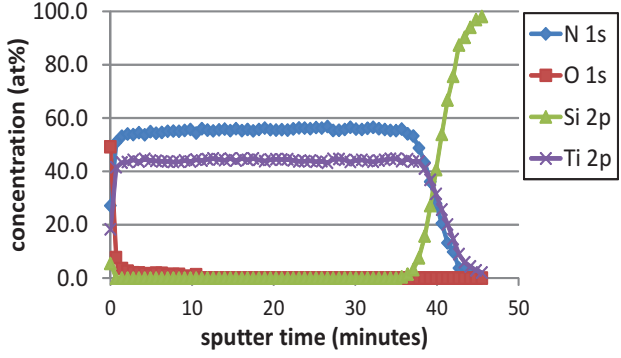}
			\caption{\label{fig:FIG1} (Color online) Concentration-depth profiles measured by XPS depth profiling. The sputter rate calibrated in $SiO_{2}$ at the used setting of the ion gun was 4.4 nm/minute. A small oxygen contamination is observed at the surface (left) on few nm, and the Ti and N concentration remain constant in the film up to the Si substrate (right). The measured stoichiometry is 1:1.3, with however an important error due to the 20$\%$ uncertainty on the absolute concentration values with this technique. 
}\end{figure}

The film has been prepared by a DC magnetron sputtering system. It is sputtered on a nitrogen/argon plasma at $350^{\circ}C$ on a high resistive HF cleaned silicon substrate (See ref. \cite{Creemer2008} for more details on the recipe). The thickness determined with a scanning electron microscope is $d=100\pm5\, nm$. The homogeneity has been checked by X-Ray Photoelectron Spectroscopy (XPS) depth profiling : as shown fig. \ref{fig:FIG1}, there is no contamination except some oxygen at the surface on few nm, and the $Ti/N$ ratio is thickness independent. The stoichiometry determined is $1:1.3$ with an uncertainty of 
20$\%$ related to the XPS calibration. There is a 1 GPa stress in the film, to obtain a dense and homogeneous granular morphology \cite{Creemer2008}. This affects the unit cell size of $4.35\,\AA$ determined by X-ray Diffraction (XRD), to be compared to the bulk value $4.24\,\AA$. XRD spectra also show a favored (200) crystalline orientation. A typical grain size of 10 nm has been determined by transmission electron microscopy in others TiN films from the same source \cite{Creemer2008}.
 
\begin{figure}[t]
		\includegraphics[width=1\linewidth]{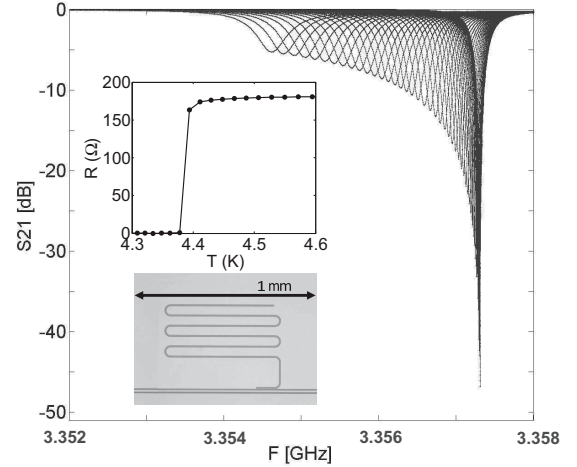}
			\caption{\label{fig:FIG2}Typical $S_{21}$ transmission of one resonator for several temperatures between $90\, mK$ and $1\, K$. When increasing T, the frequency and the quality factor decrease due to quasiparticle excitations above the gap. Upper inset: resistive superconducting transition. Lower inset: photo of one resonator from another chip having the same design. The meandered resonator line (light grey) is separated to the groundplane all around (light grey) by slits (dark grey) and is capacitively coupled to the feedline (lower straight line).
}\end{figure}

The film characterizations at low temperature include standard R(T) and Hall measurements. The resistivity at $10\,K$, $\rho=92\,\mu\Omega cm$ is almost constant up to 300 K and the carrier density is $n_{0}=3.77\,10^{22}cm^{-3}$. $k_{F}\ell$ can be estimated in the free electron model from $\rho$ and $n_{0}$ by \cite{Ashcroft} : $k_{F}\ell=3\pi^{2}\hbar/(e^{2}\rho\sqrt[3]{3\pi^{2}n_{0}})$ giving 12.7 for this film. In addition, fig.\ref{fig:FIG2} shows the sharp superconducting transition observed at $T_{C}=4.38\pm0.01\, K$. 
 
Coplanar waveguide resonators have been patterned using standard contact lithography and dry etching with an $SF_{6}/Ar$ gas mixture. One resonator is shown fig.\ref{fig:FIG2}. The resonators are formed by a central meandered line of few mm length, $3\,\mu m$ wide, and slits of $2\,\mu m$ wide between the central line and the groundplane. They are capacitively coupled to the feedline by placing one resonator end alongside it. The feedline is connected to coaxial cables at both chip sides and its $S_{21}$ transmission is measured using a standard Vector Network Analyzer. A detailed description of the setup can be found in \cite{Baselmans_JLTP12}.  
 
Each resonator gives rise to a dip in $S_{21}$ at a frequency $f_{0}^{-1}=2\pi xl\sqrt{(L_{k}+L_{g})C}$, with $l$ the resonator length, $L_{k}$ and $L_{g}$ the kinetic and geometric inductance per unit length and $C$ the capacitance per unit length. $x$ is a factor depending on the resonator type. 
Here, half of the resonators are halfwaves (open ended on both sides) corresponding to $x=2$, the others are quarterwaves (short ended on one side) thus $x=4$. 
The inductance per unit length is proportional to the surface inductance: $ L_{s}=L_{k}/g $ with g a factor depending on the resonator geometry \cite{Driessen2012}. When the temperature is increased, Cooper pairs are broken by thermal excitations above the gap, resulting in a change of the kinetic inductance. This translate into a resonance frequency shift (See fig.\ref{fig:FIG2}) and
\begin{equation}
\label{11}
\frac{\delta f\left(T\right)}{f}=-\frac{\alpha}{2}\frac{\delta L_{k}\left(T\right)}{L_{k}}=-\frac{\alpha}{2}\frac{\delta 
L_{s}\left(T\right)}{L_{s}}
\end{equation}
With $\alpha\equiv L_{k}/(L_{k}+L_{g})$ the kinetic inductance ratio. 
For high $ L_{k}\sim L_{g} $ films like the TiN film studied here, $\alpha$ is large enough to be determined precisely from the ratio of the experimental resonance frequency at the lowest temperature $f_{0}$ and the geometrical frequency $f_{g}$ calculated from the CPW dimensions \cite{Gao_NIMPR06}.

To choose the correct model for $L_{s}(T)$, we first estimate several characteristic lengths . 
The London magnetic penetration depth $ \lambda_{L}\approx47\,nm $ is determined from the measured $n_{0}$ and using a quasiparticle mass $ m^{*} $ equal to 3 times the electron mass\cite{Walker_SS98}. 
The BCS coherence length $\xi_{0}=\frac{\hbar v_{F}}{\pi\Delta}\approx120\, nm$ is estimated a priori using a gap $\Delta=1.76\,k_{B}T_{C}$ and a Fermi velocity $v_{F}=\frac{\hbar \sqrt[3]{3\pi^{2}n_{0}}}{m^{*}} $. 
The mean free path $\ell=\tau v_{F}=\frac{m^{*}}{n_{0}e^{2}\varrho}v_{F}\approx1\,nm $ is calculated from the measured $\varrho_{0}$ and $n_{0}$. We are now in position to estimate the effective coherence length $\xi_{eff}=\left(\xi_{0}^{-1}+\ell^{-1}\right)^{-1}\approx1.0\, nm$ and the magnetic penetration depth $\lambda_{eff}=\lambda_{L}\left(1+\frac{\xi_{0}}{\ell}\right)^{1/2}\approx517\, nm$. 
The film studied here is in the local ($\xi_{eff}\ll\lambda_{eff}$), dirty ($ \ell\ll\xi_{0},\lambda_{L} $), three dimensional ($ \xi_{eff}\ll d $) and thin film limit ($ d\ll\lambda_{eff} $).
In thin films, the current is distributed uniformly and the relation between the square inductance and $\lambda _{eff} $ simply reduces to: 
\begin{equation}
\label{12}
L_{s}=\mu_{0}\frac{\lambda_{eff}^{2}}{d} 
\end{equation}
In the local dirty limit and at $ T<T_{c}/3 $, the temperature dependence of $ \lambda _{eff} $ is given by \cite{TinkhamBook}:
\begin{equation}
\label{13}
\frac{\lambda_{eff}\left(T\right)}{\lambda_{eff}\left(0\right)}= tanh\left(\frac{\Delta_{0}}{2k_{B}T}\right)^{-1/2}
\end{equation} 
With $\Delta_{0}$ the superconducting gap at zero temperature. This is valid in the low frequency limit $hf\ll 2\Delta_{0}$. It is verified here, since the frequency range used is $3-5.3\, GHz$ corresponding to $12-22\,\mu eV$ which is less than 2$\%$ of the gap energy. 
Combining eq.\ref{11}, \ref{12} and \ref{13} one obtains:
\begin{equation}
\label{FitF}
\frac{\delta L_{s}}{L_{s0}}=-\frac{2}{\alpha}\frac{\delta f}{f_{0}}=2\left(tanh\left(\frac{\Delta_{0}}{2k_{B}T}\right)^{-1/2}-1\right)
\end{equation}
Where $ f $, $ L_{s} $ and $ \lambda_{eff} $ have been replaced by their value at the lowest experimental temperature $ f_{0} $, $ L_{s0} $ and $ \lambda_{eff0} $ which is correct for $ \delta L_(s)\ll L_{s0} $ and $ \delta \lambda_{eff}(T)\ll \lambda_{eff 0} $. This is valid for all temperatures below $ T\sim0.9\,T_{c} $, temperature above which the magnetic penetration depth diverges.
As shown by eq. \ref{FitF}, the temperature dependence of the resonance frequency shift is a direct probe of the superconducting gap. 

\begin{figure}[t]
		\includegraphics[width=1\linewidth]{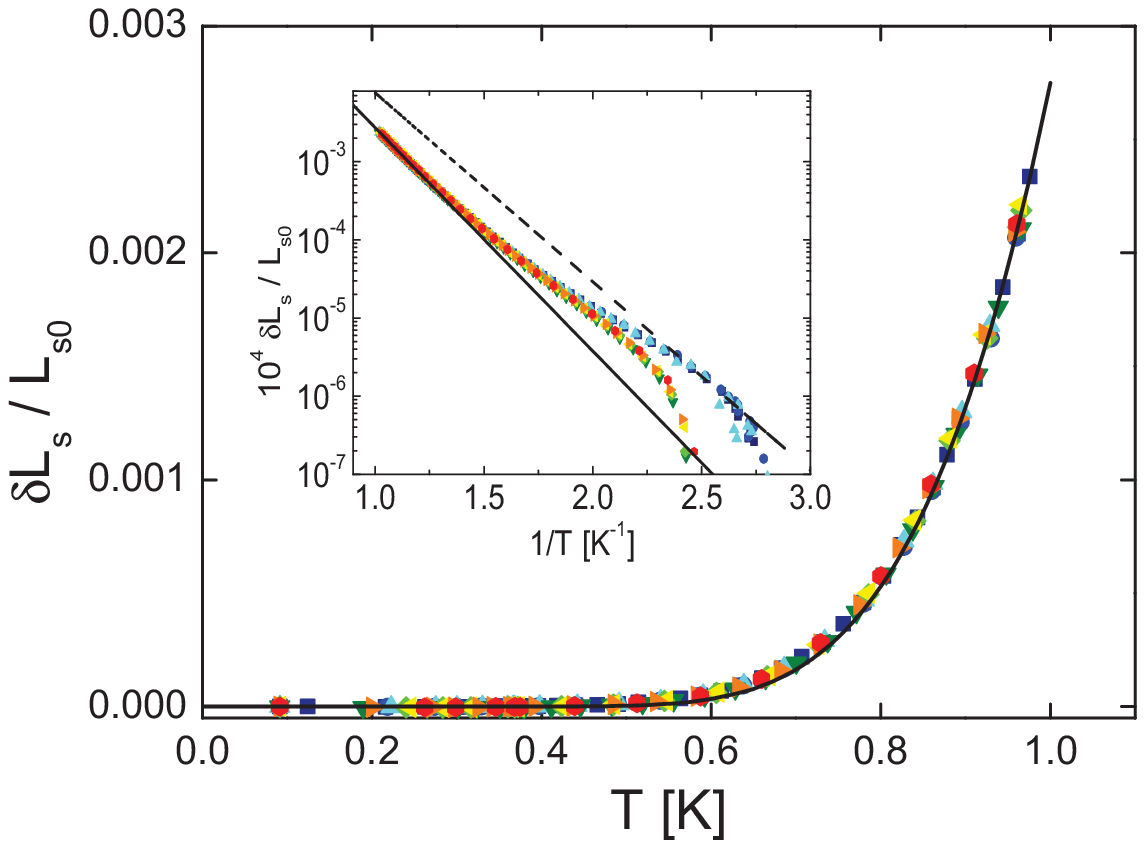}
			\caption{\label{fig:FIG3}(Color online) Temperature dependence of the normalized surface inductance for the 8 resonators, having $f_{0}=3.28\,GHz$ ($\color{Navy} \mbox\FilledSmallSquare$), $3.36\,GHz$ ($\color{blue} \mbox\FilledSmallCircle$), $3.41\,GHz$ ($\color{cyan} \mbox\FilledSmallTriangleUp$), $3.46\,GHz$ ($\color{olive} \mbox\FilledSmallTriangleDown$), $3.52\,GHz$ ($\color{green} \mbox\FilledSmallDiamondshape$), $4.79\,GHz$ ($\color{yellow} \mbox\FilledSmallTriangleLeft$), $5.01\,GHz$ ($\color{orange} \mbox\FilledSmallTriangleRight$), $5.23\,GHz$ ($\color{red} \mbox\FilledSmallCircle$). The black line is the best fit with eq. \ref{FitF} giving  the gap value $\Delta=0.57\, meV$. Inset: same curves in semi-log and inverse temperature scales to magnify the results at the lowest temperatures. The dotted line is eq. \ref{FitF} with $\Delta=0.48\, meV$.
}\end{figure}

The $S_{21}$ transmission of 8 resonators of the same chip has been measured between $90\, mK$ and $1\, K$. 
At low temperature, the internal quality factors are between $10^{6}$ and $10^{7}$. 
These high values support the conclusions of ref. \cite{Vissers_APL10} on the relation between high quality factors and a (200) crystalline orientation in TiN films. 
The $ \left(f_{0}/f_{g}\right) $ ratio calculated from the measured $ f_{0} $ is the same for all resonators, giving $\alpha=1-\left(f_{0}/f_{g}\right)^{2}=0.69$.
$\delta L_{s}/L_{s0}(T)$ is shown in fig.\ref{fig:FIG3} for all resonators. They exhibit the same temperature dependence and an excellent fit is obtained with eq.\ref{FitF}, giving $\Delta=0.57\, meV$. 
There is however a slightly weaker curvature of the measured $\delta L_{s}/L_{s0}(T)$ below $ 0.6\,K $, which is magnified in log scale in the inset. 
The curves are between the values expected for $\Delta=0.48\, meV$ and $\Delta=0.57\, meV$, which may be attributed to a small gap decrease close to the film surfaces due to oxidization or stoichiometry changes, see fig.\ref{fig:FIG1}. 
The inflexion point around $ T\sim0.45K $, followed below by a cut-off of the inductance shift which is not identical for all resonators/frequencies, is due to the capacitive dielectric variations \cite{Barends_APL08} which becomes non negligible compared to the inductive changes in this temperature range. 
Comparing $\Delta$ and the critical temperature $T_{C}=4.38\, K$, one obtains a ratio $\Delta/k_{B}T_{C}=1.50$, lower than the BCS weak coupling ratio 1.76. 
This has nothing to do with the SIT : in contrast, as observed by Sacepe et al. in TiN ultra thin films \cite{Sacepe_PRL08}, one expect disorder  to possibly increase $\Delta/k_{B}T_{C}$ close to the transition. 
The 1.50 ratio may be related to an effect of the grain size: as reported by Bose et al. \cite{Bose_PRL05} in niobium thick films, the critical temperature decreases when decreasing the grain size  and the $\Delta/k_{B}T_{C}$ ratio is slightly reduced, going to 1.61 for 18 nm grains. A superconductivity weakening can also occur due to interface tunnel exchange at internal and external surfaces\cite{Halbritter_PRB92} as observed in niobium thin films\cite{Halbritter_SSC80}. This may also explain why all $T_{C}$ reported for TiN films are below 4.8 K whereas $T_{C}=6.0K$ in a TiN single crystal \cite{Spengler_PRB78}.

\begin{figure}[t]
		\includegraphics[width=1\linewidth]{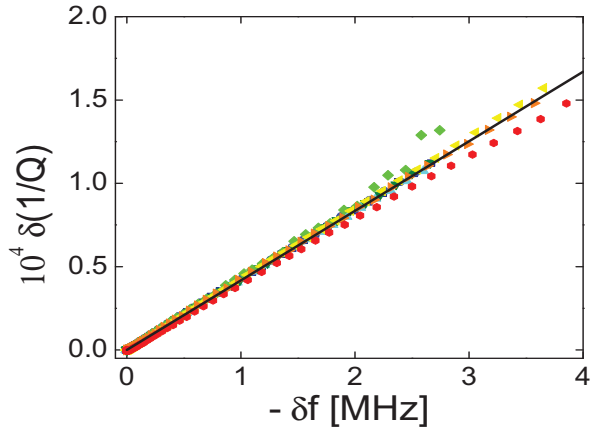}
			\caption{\label{fig:FIG4}(Color online) Inverse of the quality factor $\delta\left(1/Q\right)(T)$ as a function of the frequency shift 
$\delta f(T)$ for the 8 resonators (same symbols as fig.\ref{fig:FIG3}) between $90\,mK$ and $1\,K$. The black line is a linear fit with eq.\ref{23} with $ \tau=1.7\,ps $. 
}\end{figure}

To go further, we have also compared the temperature dependence of the frequency and the quality factor $Q$. 
Fig. \ref{fig:FIG4} shows the temperature dependence of the normalized quality factors $\delta(1/Q)\left(T\right)=\left(1/Q\right)\left(T\right)-\left(1/Q
\right)(T=0)$ for all resonators as a function of the frequency shift $-df(T)$. The 8 resonators gives similar results, and exhibit a linear behaviour. 
In the following, we show how this proportionality between quality factor and frequency shift can be reproduced analytically in the context of the two fluid model.
In general, the radio frequency absorption cannot be described by local electrodynamics even at low temperature and low frequency, due to the non trivial form of the momentum transition  matrix M \cite{Halbritter_ZPhys74}. This holds here however, since we are only interested by the temperature dependence $ \delta(1/Q)(T) $ and because M is temperature independent at $ T<T_{c}/3 $ where the gap and BCS coherence length remain unchanged.

$Q$ is related to the surface impedance 
$Z_{S}=R_{S}+i\omega L_{S}$ by 
$Q=\omega L_{total}/R_{total}=\omega L_{S}/\alpha R_{S}$, with $R_{S}$ the surface resistance.
In the thin film limit, $Z_{S}$ is simply related to the complex conductivity $\sigma=\sigma_{1}+i\sigma_{2}$ by $ Z_{s}^{-1}=\sigma d $. At low temperature, $\sigma_{1}\ll\sigma_{2}$ and we get $R_{S}=\frac{\sigma_{1}}{\sigma_{2}^{2}d}$ 
and $L_{S}=\frac{1}{\omega\sigma_{2}d}$.
The complex conductivity in the two fluid model is given by \cite{TinkhamBook}:
\begin{equation}
\sigma=\frac{n_{n}e^{2}\tau}{m^{*}}\frac{1}{1-i\omega\tau}+i\frac{n_{s}e^{2}}{m^{*}\omega}
\end{equation}
With $n_{n}$ the quasiparticle density and $\tau$ the momentum relaxation time.
In the limit $\omega\tau\ll1$ one gets for the quality factor:
\begin{equation}
\label{20}
\frac{1}{Q}=\alpha\frac{\sigma_{1}}{\sigma_{2}}=\alpha\omega\tau\frac{n_{n}}{n_{S}}
\end{equation} 
The first part of eq.\ref{20}  is identical to the equation used by Gao et al. \cite{Gao_JLTP08} in the context of the Mattis Bardeen theory. Using the properties $n_{n}=\delta n_{n}$, $\delta n_{n}/\delta n_{s}=-2$, it becomes:
\begin{equation}
\label{21}
\frac{1}{Q}(T)=-2\alpha\omega\tau\frac{\delta n_{S}(T)}{n_{S}}
\end{equation} 
The quality factor temperature dependence is directly related to the superfluid density changes, as for the kinetic inductance: $L_{S}(T)=\frac{m}{2n_{S}(T)e^{2}d}$ and
\begin{equation}
\label{22}
\frac{\delta n_{S}}{n_{S}}=-\frac{\delta L_{s}}{L_{s}}=\frac{2}{\alpha}\frac{\delta f}{f}
\end{equation} 
By combining eq.\ref{21} and eq.\ref{22} we obtain the relation between Q and $\delta f$ : 
\begin{equation}
\label{23}
\frac{1}{Q}=-8\pi\tau \,\delta f
\end{equation}

As shown by eq.\ref{23}, the inverse quality factor is proportional to the frequency shift and momentum relaxation time. 
This means that losses, which are expected to be zero at zero temperature, increase proportionally to the quasiparticle density changes $ \delta f \propto \delta n_{n} $, via the two fluid property $ \delta n_{n} \propto \delta n_{s}$. 
In practice, there are always residual losses $ 1/Q(T=0) $ due to non equilibrium excess quasiparticles \cite{Oates_PRB91, DeVisser_PRL11}, which are subtracted in fig.\ref{fig:FIG4}. 
Fitting the results with eq.\ref{23} gives the momentum relaxation time at $T\ll T_{c} $, $\tau=1.7\,10^{-12}\, s$. This is three orders of magnitude longer than in the metallic state $\tau_{m}=m^{*}/n_{0}e^{2}\rho=3.0\,10^{-15}\, s$ estimated from the resistivity and the carrier density measured at $10 K$. In the superconducting state indeed, the momentum relaxation time strongly increases, typically up to $\sim10^{-12}\, s$ in usual BCS superconductors \cite{TinkhamBook} due to the quasiparticles vanishing by Cooper pair condensation.

The temperature dependence of the resonators frequencies and quality factors clearly evidence the presence of one well defined superconducting gap in the density of states. Moreover, the excellent reproducibility between the resonators evidence the good superconducting phase homogeneity in the film. This differs from previous STM/S results on a similar $100\, nm$ thick high $T_{C}=4.68\, K$ TiN film, which report on the presence of an inhomogeneous superconducting phase having local normal areas. As discussed in ref. \cite{Escoffier_PRL04}, the detected inhomogeneous gap may be due to mesoscopic fluctuations at the proximity of a superconductor to metal transition. Indeed, the presence of a composition driven transition in TiN$_{x}$ has been reported recently\cite{Leduc_APL10}, with a disappearance of the superconducting phase in the low $x$ range, recovered at $x=0$ (titanium). However, the film used in \cite{Escoffier_PRL04} has the characteristic high critical temperature of overstoichiometric TiN, similar to the one of the film measured here, and our results doesn't exhibit any signature of the proximity with the composition transition.
Additionally, the granular morphology of the two films are different. Indeed, the film recipe used here has been especially developed to obtain densely packed grains in the film, corresponding to zone T in the Thornton classification. As discussed in ref. \cite{Creemer2008} , a typical low stress TiN film is of zone 1, containing many voids between grains. This may lead to important Josephson barriers between superconducting grains, which are expected to play a major role in the homogeneity of the superconducting phase of such systems even at relatively low disorder \cite{Spivak_PRB08}. 

In conclusion, the low frequency complex surface impedance has been studied on microresonators made from a thick overstoichiometric TiN film. The resonators high quality factors allow a high sensitivity determination of the inductive and resistive changes with temperature. In contrast to the previous spectroscopic study in a similar film, all our results are in agreement with an homogeneous superconducting phase having a gap $\Delta=0.57\, meV=1.50k_{B}T_{C}$.

We thank T. M. Klapwijk, A. Endo, E. F. C. Driessen, P. C. J. J. Coumou, R. R. Tromp and S. J. C. Yates for support and fruitful discussions.


\end{document}